\documentclass{aa}
\usepackage{natbib}
\bibpunct{(}{)}{;}{a}{}{,} 
\usepackage{graphicx}
\usepackage[varg]{txfonts}
\usepackage{bm}
\usepackage{hyperref}
\hypersetup{
    colorlinks=true,
    linkcolor=blue,
    citecolor=blue
    }
\usepackage{yhmath}
\usepackage{booktabs,multirow}
\pdfmapfile{+txfonts.map}
\begin{document} 

   \title{Methodology for estimating the magnetic Prandtl number and application to solar surface small-scale dynamo simulations}
   \titlerunning{Estimating the magnetic Prandtl number of small-scale dynamo simulations}
   \author{F. Riva
          \inst{1}
          \and
          O. Steiner\inst{1,2}
          }
    \authorrunning{F. Riva \& O. Steiner}

   \institute{Istituto Ricerche Solari Locarno (IRSOL), Università della Svizzera italiana (USI), CH-6605 Locarno-Monti, Switzerland\\
              \email{fabio.riva@irsol.usi.ch}
         \and
             Leibniz-Institut für Sonnenphysik (KIS), Schöneckstrasse 6, 79104 Freiburg i.Br., Germany
             }
             
    \date{Received xxx / Accepted yyy}

 
  \abstract
  {A crucial step in the numerical investigation of small-scale dynamos in the solar atmosphere consists of an accurate determination of the magnetic Prandtl number, $Pr_\mathrm{m}$, stemming from radiative magneto-hydrodynamic (MHD) simulations.}
   {The aims are to provide a reliable methodology for estimating the effective Reynolds and magnetic Reynolds numbers, $Re$ and $Re_\mathrm{m}$, and their ratio $Pr_\mathrm{m}=Re_\mathrm{m}/Re$ (the magnetic Prandlt number), that characterise MHD simulations and to categorise small-scale dynamo simulations in terms of these dimensionless parameters.}
   {The methodology proposed for computing $Re$ and $Re_\mathrm{m}$ is based on the method of projection on proper elements and it relies on a post-processing step carried out using higher order accurate numerical operators than the ones in the simulation code. A number of radiative MHD simulations with different effective viscosities and plasma resistivities were carried out with the CO\textsuperscript{5}BOLD code, and the resulting growth rate of the magnetic energy and saturated magnetic field strengths were characterised in terms of $Re$ and $Re_\mathrm{m}$.}
   {Overall, the proposed methodology provides a solid estimate of the dissipation coefficients affecting the momentum and induction equations of MHD simulation codes, and consequently also a reliable evaluation of the magnetic Prandtl number characterising the numerical results. Additionally, it is found that small-scale dynamos are active and can amplify a small seed magnetic field up to significant values in CO\textsuperscript{5}BOLD simulations with a grid spacing smaller than $h=12\ \mathrm{km}$, even at $Pr_\mathrm{m}\simeq0.65$. However, it is also evident that it is difficult to categorise dynamo simulations in terms of $Pr_\mathrm{m}$ alone, because it is not only important to estimate the amplitude of the dissipation coefficients, but also at which scales energy dissipation takes place.}
   {}


   \maketitle
%

\section{Introduction}
    Over the past two decades, leveraging on the work done by Nordlund in the 1980s~\citep{Nordlund1982}, radiative magneto-hydrodynamic (MHD) simulations took on an increasingly important role in improving our understanding of the non-linear processes observed on the Sun. Unfortunately, the discretisation of an MHD model and its implementation in a simulation code always introduce numerical errors in the system, which in turn add numerical viscosity and magnetic diffusivity to the model equations~(see, e.g. \citealt{laney_1998,bodenheimer2006numerical,Margolin2019}). This poses two critical problems. First, it is currently impossible to simulate the solar atmosphere at realistic Reynolds and magnetic Reynolds numbers, $Re=LU/\nu$ and $Re_\mathrm{m}=LU/\eta$, with $L$ and $U$ the characteristic length scale and velocity of the plasma flow and $\nu$ and $\eta$ the molecular viscosity and magnetic diffusivity (the latter also known as plasma resistivity). The reason is that the plasma flows in the solar atmosphere are characterised by large Reynolds and magnetic Reynolds numbers and by very small magnetic Prandtl numbers, $Pr_\mathrm{m}=Re_\mathrm{m}/Re=\nu/\eta$, with $Pr_\mathrm{m}$ in the $10^{-7}-10^{-4}$ range depending on the depth~\citep[see, e.g.][]{Brandenburg2005,Schekochihin2005}. Therefore, it is necessary to resolve the wide range of length scales $L\gg l_\eta\gg l_\nu$ to simulate the solar atmosphere reliably~\citep{Schekochihin2007}. Here, $l_\nu$ and $l_\eta$ are the viscous and resistive length scales, respectively. The second issue is related to the fact that the intrinsic diffusivities of a simulation code complicate the interpretation of the results obtained with it, since the effective Reynolds and magnetic Reynolds numbers stemming from a simulation, $Re_\mathrm{eff}$ and $Re_{\mathrm{m},\mathrm{eff}}$, are generally unknown.
    
    One of the major open questions in solar physics concerns the origin of the ubiquitous small-scale magnetic field observed in the solar photosphere~\citep[see, e.g.][]{2011ASPC..437..451S}. Since polarimetric signals measured in the inter-network of the quiet Sun are at the limit of present telescope capabilities and different observations of different resolutions and measurement techniques provide different results, it is unclear what the exact mean strength and spatial and angular distribution of the magnetic field in the solar atmosphere~are \citep[see, e.g.][and references therein]{MartinezPillet2013,Khomenko2017}. Given these uncertainties, it is difficult to understand if the small-scale magnetic field of the quiet Sun mainly originates from a turbulent cascade acting on fields produced by a global-scale dynamo, or if it is mainly generated locally by a small-scale turbulent dynamo, as first suggested by Petrovay and Szakaly in \citeyear{PS1993}. The fact that properties of the quiet Sun's magnetic field seem independent of the solar cycle~\citep{Trujillo2004,Almeida2004,Buehler2013,Lites2014,Ramelli2019} would favour the latter explanation. However, hints of contradictory results were obtained by Kleint et al. in~\citeyear{Kleint2010}.
    
    To shed light on this controversy, a number of numerical studies have been carried out in the past two decades. Initially, the possibility of driving dynamo action in the quiet photosphere through turbulent convection was demonstrated numerically by~\citet{Cattaneo1999}. Later studies of small-scale dynamos in the solar atmosphere were performed with realistic solar-like numerical simulations by~\citet{Vogler2007}, showing that a considerable amount of magnetic energy in the solar photosphere could be self-sustained through dynamo action. More recently, the possibility of amplifying a small seeded magnetic field ($\sim10^{-6}-10^{-2}\ \mathrm{G}$) in the solar photosphere through dynamo action was investigated by means of radiative MHD simulations in a number of additional publications~\citep[see, e.g.][]{Graham2010,Rempel2014,Kitiashvili2015,Thaler2015,Khomenko2017}.
    
    Despite these numerical studies providing extremely useful information on many aspects of small-scale dynamo action in turbulent plasma flows, it is still unclear how to extrapolate these results to the Sun. This is for two reasons. First, the classical picture discussed in~\citet{Batchelor1950} and generally used to explain field amplification at $Pr_\mathrm{m}\gtrsim1$, the regime typically accessible by current small-scale solar dynamo simulations, does not apply to the solar atmosphere. This is because the fundamental assumption that the scale of the fluid motion that does the stretching is larger than the scale of the field that is stretched, is not valid for $Pr_\mathrm{m}\ll1$~\citep{Schekochihin2007}. In this context, it is still unclear if $\lim_{Pr_\mathrm{m}\rightarrow0}Re_{\mathrm{m},\mathrm{c}}=\infty$, with $Re_{\mathrm{m},\mathrm{c}}$ the critical magnetic Reynolds number for dynamo action, or if instead $Re_{\mathrm{m},\mathrm{c}}$ saturates to some finite value in the limit $Pr_\mathrm{m}\rightarrow0$. Second, the saturated averaged magnetic field resulting from small-scale dynamo simulations appears to strongly depend on the details of the simulation setup, and in particular on the effective magnetic Prandtl number, $Pr_{\mathrm{m},\mathrm{eff}}$, which is generally unknown.
    
    A crucial step in the study of small-scale dynamos in the solar atmosphere is therefore a reliable evaluation of $Pr_{\mathrm{m},\mathrm{eff}}$ and, consequently, of the effective dissipation coefficients stemming from a radiative MHD simulation,  $\nu_\mathrm{eff}$ and $\eta_\mathrm{eff}$. In this respect, a common approach consists of adding explicit, artificial diffusion terms to the model equations and assuming that these terms are much larger than the intrinsic diffusivities of the simulation code. Several methodologies have also been proposed in the recent past for estimating the effective diffusivities stemming from an MHD simulation, or at least the resulting magnetic Prandtl number. These include: simulating simple one- or two-dimensional problems for which we know the exact solution of the resistive-viscous MHD equations and estimating the value of $\nu_\mathrm{eff}$ and $\eta_\mathrm{eff}$ by fitting the results~\citep{Rembiasz2017}; adding forcing terms to the model equations and associating the resulting saturated magnetic field to a turbulent magnetic diffusivity~\citep{Fromang2009}; and relating $Pr_{\mathrm{m},\mathrm{eff}}$ to the ratio of the velocity and magnetic energy Taylor microscales by means of a homogeneous turbulence simulation~\citep{Graham2010}. However, while these procedures provide reliable results under some particular conditions, they are difficult to generalise to different flow regimes and simulation setups. This motivates the present work, with a twofold objective. First, we propose a general and rigorous methodology for estimating $\nu_\mathrm{eff}$ and $\eta_\mathrm{eff}$, and, consequently, $Pr_{\mathrm{m},\mathrm{eff}}$. Second, we investigate how the magnetic field resulting from small-scale solar dynamo simulations depends on $Re_\mathrm{eff}$ and $Re_{\mathrm{m},\mathrm{eff}}$.
    
    Our investigation is based on the method of projection on proper elements (PoPe), first introduced in~\citet{Cartier-Michaud2016} and later extended to the independent PoPe~\citep{Cartier-Michaud2020}, which has been used in the plasma physics community to quantify the error affecting numerical results and to verify plasma turbulence simulation codes. The estimate of the numerical error relies on using different, higher order accuracy numerical operators than the ones in the simulation code to post-process the simulation results. Then, we assume that the resulting numerical error can be, at least partially, modelled with a diffusion operator, and we compute the diffusivities through a least-squares fit. The proposed methodology is  applied to a number of small-scale dynamo simulations carried out with the CO\textsuperscript{5}BOLD code~\citep{Freytag2012}, with the aim of characterising the magnetic energy growth rate and the saturated magnetic field in terms of $Re_\mathrm{eff}$ and $Re_{\mathrm{m},\mathrm{eff}}$.
    
    The remainder of this paper is structured as follows. In Sect.~\ref{sec:methodology}, we review the method of PoPe and we discuss how to extend it to compute the effective dissipation coefficients stemming from an MHD simulation and the resulting magnetic Prandtl number. In Sect.~\ref{sec:setup}, we illustrate the numerical setup used for the simulations discussed in the present paper. In Sect.~\ref{sec:res_etanu}, we apply the methodology proposed in Sect.~\ref{sec:methodology} to a number of CO\textsuperscript{5}BOLD simulations, and we discuss the results. Small-scale dynamo simulations and their characterisation in terms of $Re_\mathrm{eff}$ and $Re_{\mathrm{m},\mathrm{eff}}$ are the subjects of Sect.~\ref{sec:dynamo}. Finally, in Sect.~\ref{sec:conclusions}, we report our conclusions.

\section{Overview of the methodology}\label{sec:methodology}
    The methodology we propose for estimating the effective viscosity and magnetic diffusivity stemming from a radiative MHD simulation is based on the PoPe method recently detailed in~\citet{Cartier-Michaud2016}, which is briefly summarised here for completeness. To illustrate the main steps of the procedure, we consider the induction equation for an ideal plasma,
    \begin{equation}\label{eq:induction}
        \partial_t\vec{B}=\nabla\times\left(\vec{v}\times\vec{B}\right),
    \end{equation}
    as an example, with $\vec{v}$ and $\vec{B}$ being the fluid velocity and the magnetic field, respectively. We note that $\vec{v}$ and $\vec{B}$ are functions of time and space, but this dependence is omitted in Eq.~(\ref{eq:induction}) to lighten the notation.
    
    We  rewrite Eq.~(\ref{eq:induction}) in the following more abstract form:
    \begin{equation}\label{eq:induction_th}
        \partial_t\vec{B}=\sum_{i=1}^2w_iO_i(\vec{v},\vec{B}),
    \end{equation}
    where 
    \begin{equation}\label{eq:wOi}
        \{w_i\}=\{1,-1\}\ \mathrm{and}\  \{O_i(\vec{v},\vec{B})\}=\{\nabla\cdot(\vec{B}\vec{v}),\nabla\cdot(\vec{v}\vec{B})\}
    \end{equation}
    are the sequences of weights and operators, respectively, and $i=1,2$. This can be considered as the projection of the time derivative of $\vec{B}$ on the basis of elements $\{O_i(\vec{v},\vec{B})\}$ with weights $\{w_i\}$. Moreover, we denote the magnetic field resulting from a simulation code implementing the discretised version of Eq.~(\ref{eq:induction}) of order of accuracy $p$ and obtained on a mesh with degree of refinement $h$ as $\vec{B}^h$, such that $\epsilon^h=\|\vec{B}-\vec{B}^h\|=C\cdot h^p+\mathcal{O}(h^{p+1})$, where $\epsilon^h$ is the numerical error affecting $\vec{B}^h$, $C$ is a constant independent of $h$, $\|\cdot\|$ is a designed norm, and $\mathcal{O}(h^{p+1})$ denotes a term that decreases to zero as $h^{p+1}$ in the asymptotic regime. The same notation is used for the fluid velocity, where the numerical solution is denoted as $\vec{v}^h$. We also assume that we have a post-processing scheme of order of accuracy $q\geq p$, where the discretised counterparts of $\partial_t(\cdot)$ and $\{O_i(\cdot,\cdot)\}$ are denoted as $\delta_t^{\mathrm{pp},h}$ and $\{O_i^{\mathrm{pp},h}(\cdot,\cdot)\}$, respectively.
    We note that, in general, if the post-processing scheme does not correspond exactly to the numerical algorithm implemented in the simulation code used to obtain $\vec{B}^h$ and $\vec{v}^h$,
    $\|\delta_t^{\mathrm{pp},h}\vec{B}^h-\sum_{i=1}^2w_iO_i^{\mathrm{pp},h}(\vec{v}^h,\vec{B}^h)\|\neq0$.
    
    The next step consists of writing
    \begin{equation}\label{eq:dBdtpp}
        \bigg\lVert\delta_t^{\mathrm{pp},h}\vec{B}^h-\sum_{i=1}^2\left[(w_i+\delta w_i)O_i^{\mathrm{pp},h}(\vec{v}^h,\vec{B}^h)\right]\bigg\rVert_2=r^h(\{\delta w_i\}),
    \end{equation}
    where $r^h(\{\delta w_i\})$ is the residual error and the $\delta w_i$ are interpreted as changes of the weights from $w_i$ to $w_i+\delta w_i$ due to the numerical error introduced by the simulation code. The $\delta w_i$ can be estimated by minimising the residual error over a set of $N\geq2$ space-time points $\vec{x}_1,...,\vec{x}_N$. In matrix form, this linear least-squares regression writes as
    \begin{equation}\label{eq:matrix}
        A^TA\bm{\tilde{w}}=A^T\vec{b},
    \end{equation}
    where $A$ is a matrix with $N\times2$ elements $a_{j,i}=O_i^{\mathrm{pp},h}[\vec{v}^h(\vec{x}_j),\vec{B}^h(\vec{x}_j)]$, $\bm{\tilde{w}}$ is a column vector with two elements $\tilde{w}_i=w_i+\delta w_i$, and $\vec{b}$ is a column vector with $N$ elements $b_j=\delta_t^{\mathrm{pp},h}\vec{B}^h(\vec{x}_j)$. The PoPe methodology states that, if a code is bug free, the expected value of $\tilde{w}_i$ converges to $w_i$, that is, the $\delta w_i$ vanish at rate $p$ in the limit $h\rightarrow0$.
    
    To proceed further, we note that if $q>p$ and $h$ is sufficiently small, the numerical error introduced by the post-processing scheme is negligible with respect to the numerical error introduced by the simulation code used to obtain $\vec{B}^h$ and $\vec{v}^h$. Moreover, assuming that a non-negligible part of the numerical error can be modelled with a diffusion term of the form $\eta_\mathrm{eff}\nabla^2\vec{B}$, we can extend the list of operators $\{O_i^{\mathrm{pp},h}(\cdot,\cdot)\}$ to include a discretised counterpart of $\nabla^2\vec{B}$, $(\nabla^2)^{\mathrm{pp},h}\vec{B}^h$ and project the residual error on it to compute the effective magnetic diffusivity $\eta_\mathrm{eff}$. In the following, we consider two different methods for performing this projection. The first method, denoted in the following as method (i), consists of minimising 
    \begin{equation}\label{eq:i}
        \bigg\lVert\delta_t^{\mathrm{pp},h}\vec{B}^h-\sum_{i=1}^2[\tilde{w}_iO_i^{\mathrm{pp},h}(\vec{v}^h,\vec{B}^h)]-\eta_\mathrm{eff}(\nabla^2)^{\mathrm{pp},h}\vec{B}^h\bigg\rVert_2
    \end{equation}
    to obtain $\tilde{w}_1,\tilde{w}_2$, and $\eta_\mathrm{eff}$. This leads to adding a column to the matrix $A$ with elements $a_{j,3}=(\nabla^2)^{\mathrm{pp},h}\vec{B}^h(\vec{x}_j)$ and a row to the vector $\bm{\tilde{w}}$ with element $\tilde{w}_3=\eta_\mathrm{eff}$ and solving Eq.~(\ref{eq:matrix}) for $\tilde{w}_1,\tilde{w}_2$, and $\eta_\mathrm{eff}$ over a set of $N\geq3$ space-time grid points. The second method, (ii), assumes that $\delta w_i=0$, and requires minimising the distance
    \begin{equation}\label{eq:ii}
        \bigg\lVert\delta_t^{\mathrm{pp},h}\vec{B}^h-\sum_{i=1}^2[w_iO_i^{\mathrm{pp},h}(\vec{v}^h,\vec{B}^h)]-\eta_\mathrm{eff}(\nabla^2)^{\mathrm{pp},h}\vec{B}^h\bigg\rVert_2
    \end{equation}
    over a set of $N\geq1$ space-time grid points to obtain $\eta_\mathrm{eff}$. In other words, the first approach consists of splitting the numerical error in two sources, one that modifies the weights from $w_i$ to $\tilde{w}_i$ and another that introduces numerical dissipation into the induction equation, whereas the second method assumes that the weights are unchanged and the numerical errors just increase the diffusion of the scheme. We note that the second approach can be understood as a more conservative estimate of the intrinsic magnetic diffusion and shares some similarities with the approach discussed in~\citet{Khomenko2017} for the evaluation of $\eta$, and it was used in~\citet{Cartier-Michaud2020} for estimating the effective diffusion affecting the continuity equation in the TOKAM3X code. We also note that, in general, we can solve Eq.~(\ref{eq:matrix}) multiple times, each time considering a few different grid points, and therefore obtaining a statistical distribution of $\eta_\mathrm{eff}$, or, we can solve Eq.~(\ref{eq:matrix}) considering a large number of points at once, thus reducing the statistical error in the estimate of $\eta_\mathrm{eff}$.
    
    A similar procedure can be used to estimate the intrinsic kinematic viscosity affecting the momentum equation. More precisely, the momentum equation is written in a form similar to Eq.~(\ref{eq:induction_th}) as $\partial_t(\rho\vec{v})=\sum_{i=1}^4w_iO_i(\rho,\vec{v},\vec{B})$, where the lists of weights and operators are now 
    \begin{equation}\label{eq:wOm}
    \begin{split}
        &\{w_i\}=\{-1,1,-1,1\}\ \mathrm{and}\\ &\{O_i(\rho,\vec{v},\vec{B})\}=\{\nabla\cdot(\rho\vec{v}\vec{v}),\nabla\cdot(\rho\vec{B}\vec{B})/(4\pi),\nabla[P+B^2/(8\pi)],\vec{g}\},
        \end{split}
    \end{equation}
     respectively, and $i=1,2,3,4$, with $\rho$ being the mass density, $P$ the plasma pressure, and $\vec{g}$ the gravity field. A post-processing scheme is then used to compute the discretised counterparts of these operators and build a matrix with $N\times4$ elements, with the rows corresponding to different space-time points and the columns to different operators. The same steps done to obtain $\eta_\mathrm{eff}$ are then repeated to obtain the effective kinematic viscosity, $\nu_\mathrm{eff}$; the only difference is the form of the diffusion operator being added to the list $\{O_i(\rho,\vec{v},\vec{B})\}$, which is denoted in the following as $\nabla\cdot\bm{\tau}$, where $\bm{\tau}$ is the stress tensor, which depends on $\nu$. Finally, the effective magnetic Prandtl number stemming from an MHD simulation is computed as $Pr_{\mathrm{m},\mathrm{eff}}=\nu_\mathrm{eff}/\eta_\mathrm{eff}$. More details on the explicit form of $\bm{\tau}$, as well as a validation of the present methodology, are discussed in Sect.~\ref{sec:res_etanu}.
    
\section{Numerical setup}\label{sec:setup}
    The simulations discussed in the present paper were obtained with the CO\textsuperscript{5}BOLD code, which is designed to simulate solar and stellar atmospheres, as well as their interior, by solving the time-dependent ideal MHD equations:
    \begin{align}
        \frac{\partial\rho}{\partial t}+\nabla\cdot(\rho\mathbf{v})=&0,\\
        \frac{\partial\rho\mathbf{v}}{\partial t}+\nabla\cdot\left[\rho\mathbf{v}\mathbf{v}+\left(P+\frac{B^2}{2}\right)\mathbf{I}-\mathbf{B}\mathbf{B}\right]=&\rho\mathbf{g},\\
        \frac{\partial\mathbf{B}}{\partial t}+\nabla\cdot(\mathbf{v}\mathbf{B}-\mathbf{B}\mathbf{v})=&0,\\
        \frac{\partial\rho e_\mathrm{tot}}{\partial t}+\nabla\cdot\left[\left(\rho e_\mathrm{tot}+P+\frac{B^2}{2}\right)\mathbf{v}-(\mathbf{v}\cdot\mathbf{B})\mathbf{B}+\mathbf{F}_\mathrm{rad}\right]=&0,
    \end{align}
    in combination with a non-grey radiative transfer equation and an equation of state (eos) that connects the mass density $\rho$ and the gas pressure $P$ to the internal energy $e_\mathrm{i}$. Here, $e_\mathrm{tot}=\rho e_\mathrm{i}+\rho v^2/2+B^2/2+\rho\Phi$ is the total energy, $\Phi$ the gravitational potential, and $\mathbf{F}_\mathrm{rad}$ the frequency-integrated radiative energy flux vector. $\mathbf{I}$ is the identity matrix. The code can be used to carry out both global simulations of an entire star from the central core to the visible stellar surface, or at least a shell encompassing the convection zone, and local simulations of a small partial volume of the star under investigation, typically encompassing the surface layers where energy transport changes from predominantly convective to purely radiative.
    
    The (magneto-)hydrodynamic equations are evolved using modern flux-conserving Riemann type solvers, with a Roe solver~\citep*{Roe1986} for non-magnetic simulations, or an HLL solver~\citep*{HLL1983} that can be used both for the magnetic and the non-magnetic cases. In regions of small plasma-$\beta$, with $\beta$ the ratio of the thermal to the magnetic pressure of the plasma, the equation of the thermal energy is used instead of the equation of the total energy to avoid negative gas pressure, this at the expense of strict energy conservation. Additionally, to avoid very small time steps when the Alfvén speed ($v_A$) is high, $v_A$ can be limited by artificially reducing the strength of the Lorentz force. To ensure a divergence-free magnetic field, the constrained-transport method of~\citet{Evans1988} is used. Furthermore, explicit artificial viscous and resistive terms can be added to the momentum and induction equations to increase the dissipation of the numerical scheme. In particular, in the following we consider two different forms of artificial viscosity and one of artificial magnetic diffusivity: a homogeneous viscous coefficient, $\nu_\mathrm{H}$; a turbulent sub-grid-scale viscosity as given by the Smagorinsky model~\citep{Smagorinsky1963}, where we denote as $C_\mathrm{S}$ the Smagorinsky coefficient; and a homogeneous magnetic diffusivity $\eta_\mathrm{H}$.
    
    Concerning the radiative transfer equation, this can be solved using either a short- or a long-characteristic scheme. Additionally, to reduce the computational cost of a simulation, the diffusion approximation can be used to model the radiative transport in the deep, optically thick layers of the numerical domain. Pre-tabulated values as functions of density and internal energy, accounting for the ionisation balance of hydrogen, helium, and a representative metal, are used to solve the eos and compute the plasma pressure and temperature.
    
    The code is written in a modular form using Fortran 90 (and some Fortran 77) language, and it is parallelised using OpenMP directives. To carry out the simulations discussed in the present paper, the parallelisations of the MHD and of the long-characteristic transfer modules were extended to a hybrid, Message Passing Interface (MPI) and OpenMP, parallelisation. For additional details on CO\textsuperscript{5}BOLD, we refer the reader to~\citet{Freytag2012}.
    
    For the present paper, we considered only local simulations (box-in-a-star setup) with a Cartesian computational domain encompassing a volume of $6.0\times6.0\times2.4\ \mathrm{Mm}$, with approximately $1.6\ \mathrm{Mm}$ below and $0.8\ \mathrm{Mm}$ above the mean Rosseland optical depth $\tau_R=1$. The transfer equation was solved with the long-characteristic radiation transport module, using Rosseland mean opacities (grey opacity), and employing the diffusion approximation for approximately the first $1\ \mathrm{Mm}$ from the bottom boundary of the domain. To compute the fluxes for the HLL solver, we used a FRweno reconstruction scheme~\citep{Freytag2013}. Moreover, the gravitational field, $\vec{g}$, was vertical and uniform with a constant value of $g=27500\ \mathrm{cm\,s^{-2}}$. The thermal energy equation was activated for $\beta<0.1$ and $v_A$ was limited at a maximum of $40\ \mathrm{km\,s^{-1}}$. We used test calculations setting the limit to $90\ \mathrm{km\,s^{-1}}$ to verify that limiting $v_A$ has only a negligible impact on the results discussed in Sects.~\ref{sec:res_etanu} and \ref{sec:dynamo}.\footnote{In particular for the dynamo runs in Sect.~\ref{sec:dynamo}, this limit becomes active in the high photosphere close to the top boundary only.} The equations were then evolved in time with an Hancock predictor step.
    
    The side boundaries were periodic in both horizontal directions. The top boundary was open for fluid flows and outward radiation, with the density decreasing exponentially in the boundary (ghost) cells outside the domain, whereas magnetic fields were forced to be vertical. Moreover, to avoid reflection of acoustic waves at the top boundary, we enforced $\partial_t\varv_z=-c_\mathrm{s}\partial_z\varv_z$ and $\partial_z\vec{v}_\mathrm{h}=0$, with $c_\mathrm{s}$ being the local sound speed and $\vec{v}_\mathrm{h}$ and $\varv_z$ the horizontal and vertical velocities, respectively. At the bottom we have set an open boundary condition for fluid variables, enforcing $\partial_z\vec{v}=0$, vanishing horizontally averaged vertical mass flux, and a control on the specific entropy of the inflowing material that ensures that the model has a resulting effective temperature close to $5770\ \mathrm{K}$. The two parameters $\mathrm{C_{sChange}}=0.1$ and $\mathrm{C_{pChange}}=0.3$ were used to reduce deviations of the entropy and the pressure from the horizontal means, as detailed in~\citet{Freytag2012}. Similarly, a damping of the vertical velocity was introduced at the bottom boundary in the following form: 
    \begin{equation}
        v_z^{\mathrm{new}}=v_z-v_z\min\left[\frac{\Delta t}{t_{\mathrm{char}}}\left(C_{\mathrm{lin}}+C_{\mathrm{sqrt}}\frac{|v_z|}{\langle v_z^2\rangle_\mathrm{h}^{1/2}+\epsilon}\right),1\right],
    \end{equation}
    where $\Delta t$ is the internal time step, $t_{\mathrm{char}}=\Delta z/\langle v_\mathrm{f}+|v_z|\rangle_\mathrm{h}$ a characteristic timescale with $\Delta z$ being the grid spacing in the vertical direction and $v_\mathrm{f}$ the flux velocity as obtained from the HLL scheme, and $\epsilon$ a tiny computer number, which is of order $10^{-300}\ \mathrm{cm}\,\mathrm{s}^{-1}$ in double precision format, introduced to avoid a vanishing denominator. The notation $\langle-\rangle_\mathrm{h}$ denotes a horizontal average, and $C_{\mathrm{lin}}$ and $C_{\mathrm{sqrt}}$ are two input parameters (in the following $C_{\mathrm{lin}}=C_{\mathrm{sqrt}}=0$ if not specified otherwise). Moreover, we have set $\vec{B}=0$ in inflow regions at the bottom boundary, ensuring that no magnetic energy enters the domain at these locations. With regard to the small-scale dynamo simulations of Sect.~\ref{sec:dynamo}, this is a very conservative condition, because it allows for the removal of magnetic energy without replenishing it from outside of the computational domain. More details on the boundary conditions used in CO\textsuperscript{5}BOLD are given in~\citet{Freytag2017}.
 
    \begin{table*}\centering
    \caption{Overview of simulations with an initial uniform magnetic field $B_z=100\ \mathrm{G}$.}
    \label{tab:100G}
    \begin{tabular*}{\linewidth}{l@{\extracolsep{\fill}}cccc}
    \hline\hline
    \noalign{\smallskip}
    Simulation name & $h\ [\mathrm{km}]$ & $\nu_\mathrm{H}\ [10^{10}\ \mathrm{cm^2\,s^{-1}}]$ & $C_\mathrm{S}$ & $\eta_\mathrm{H}\ [10^{10}\ \mathrm{cm^2\,s^{-1}}]$\\
    \noalign{\smallskip}
    \hline
    \noalign{\smallskip}
    $h24$ & $24$ & $0$ & $0$ & $0$\\
    $h16$ & $16$ & $0$ & $0$ & $0$\\
    $h12$ & $12$ & $0$ & $0$ & $0$\\
    $h8$ & $8$ & $0$ & $0$ & $0$\\
    $h8\nu_\mathrm{H}4$ & $8$ & $4$ & $0$ & $0$\\
    $h8\nu_\mathrm{H}8$ & $8$ & $8$ & $0$ & $0$\\
    $h8\nu_\mathrm{H}12$ & $8$ & $12$ & $0$ & $0$\\
    $h8\nu_\mathrm{H}16$ & $8$ & $16$ & $0$ & $0$\\
    $h8\eta_\mathrm{H}4$ & $8$ & $0$ & $0$ & $4$\\
    $h8\eta_\mathrm{H}8$ & $8$ & $0$ & $0$ & $8$\\
    $h8\eta_\mathrm{H}12$ & $8$ & $0$ & $0$ & $12$\\
    $h8\eta_\mathrm{H}16$ & $8$ & $0$ & $0$ & $16$\\
    $h6$ & $6$ & $0$ & $0$ & $0$\\
    $h6\nu_\mathrm{H}$ & $6$ & $4.1$ & $0$ & $0$\\
    $h6\nu_\mathrm{S}$ & $6$ & $0$ & $1.55$ & $0$\\
    $h6\eta_\mathrm{H}$ & $6$ & $0$ & $0$ & $3.5$\\
    \noalign{\smallskip}
    \hline
    \end{tabular*}
    \tablefoot{Name, grid spacing $h$, artificial homogeneous viscosity $\nu_\mathrm{H}$, Smagorinsky coefficient $C_\mathrm{S}$, and artificial homogeneous magnetic diffusivity $\eta_\mathrm{H}$.}
    \end{table*}
 
    In this work, we considered the five grid resolutions $h=24,16,12,8$, and $6\ \mathrm{km}$, where $h$ denotes a uniform grid spacing in the three spatial coordinates. We started from a previous non-magnetic CO\textsuperscript{5}BOLD simulation of the solar atmosphere, adapted it to a computational domain encompassing a volume of $6.0\times6.0\times2.4\ \mathrm{Mm}$, interpolated the results to a grid resolution of $h=24\ \mathrm{km}$, and evolved the resulting plasma fields for more than $10\,000\ \mathrm{s}$ to damp out fluctuations introduced by this interpolation. Then, for each finer resolution, we started from a relaxed, coarser simulation, we interpolated the numerical results to the new grid, and we carried out a non-magnetic run for at least $10^4\ \mathrm{s}$. From the final state of the five relaxed non-magnetic simulations, we started two sets of magnetic simulations, with an initial uniform vertical magnetic field $B_z=100\ \mathrm{G}$ and $B_z=1\ \mathrm{mG}$.
    
    The first set of simulations (initial vertical magnetic field $B_z=100\ \mathrm{G}$) is used in Sect.~\ref{sec:res_etanu} to investigate the methodology presented in Sect.~\ref{sec:methodology}. To this aim, we carried out a total of 16 simulations, each for $1000\ \mathrm{s}$: five reference simulations with $h=24,16,12,8$, and $6\ \mathrm{km}$; four simulations with $h=8\ \mathrm{km}$ and $\nu_\mathrm{H}=4,8,12$, and $16\cdot10^{10}\ \mathrm{cm^2\,s^{-1}}$; four simulations with $h=8\ \mathrm{km}$ and $\eta_\mathrm{H}=4,8,12$, and $16\cdot10^{10}\ \mathrm{cm^2\,s^{-1}}$; one simulation with $h=6\ \mathrm{km}$ and $\nu_\mathrm{H}=4.1\cdot10^{10}\ \mathrm{cm^2\,s^{-1}}$; one simulation with $h=6\ \mathrm{km}$ and $C_\mathrm{S}=1.55$; and one simulation with $h=6\ \mathrm{km}$ and $\eta_\mathrm{H}=3.5\cdot10^{10}\ \mathrm{cm^2\,s^{-1}}$. A summary of the numerical parameters for the first set of simulations is presented in Table~\ref{tab:100G}.
    
    \begin{table*}\centering
    \caption{Overview of simulations with an initial uniform magnetic field $B_z=1\ \mathrm{mG}$.}
    \label{tab:1mG}
    \begin{tabular*}{\linewidth}{l@{\extracolsep{\fill}}ccccccccc}
    \hline\hline
    \noalign{\smallskip}
    Simulation name & $h\ [\mathrm{km}]$ & $C_\mathrm{S}$ & $\eta_\mathrm{H}\ [10^{9}\ \mathrm{cm^2\,s^{-1}}]$ & $Re_\mathrm{eff}$ & $Re_{\mathrm{m},\mathrm{eff}}$ & $Pr_{\mathrm{m},\mathrm{eff}}$ & $\tau_E\ [\mathrm{s}]$ & $\langle|B_z|\rangle\ [\mathrm{G}]$ & $\langle E_\mathrm{m}/E_\mathrm{k}\rangle\ [\%]$\\
    \noalign{\smallskip}
    \hline
    \noalign{\smallskip}
    d$h24$ & $24$ & $0$ & $0$ & 250 & 238 & 0.95 & ... & ... & ...\\
    d$h16$ & $16$ & $0$ & $0$ & 416 & 389 & 0.93 & ... & ... & ...\\
    d$h12$ & $12$ & $0$ & $0$ & 595 & 553 & 0.93 & 9200 & 9 & 0.2\\
    d$h8$ & $8$ & $0$ & $0$ & 1023 & 923 & 0.90 & 1440 & 25 & 1.3\\
    d$h6$ & $6$ & $0$ & $0$ & 1432 & 1278 & 0.89 & 830 & 49 & 1.4\\
    d$h6\nu_\mathrm{S}\eta_\mathrm{H}22$ & $6$ & $1.55$ & $22$ & 581 & 930 & 1.6 &  5200 & ... & ...\\
    d$h6\nu_\mathrm{S}\eta_\mathrm{H}25$ & $6$ & $1.55$ & $25$ & 574 & 860 & 1.5 & 12000 & ... & ...\\
    d$h6\nu_\mathrm{S}\eta_\mathrm{H}32$ & $6$ & $1.55$ & $32$ & 570 & 740 & 1.3 &  ... & ... & ...\\
    d$h6\eta_\mathrm{H}12$ & $6$ & $0$ & $12$ & 1416 & 957 & 0.68 &  5500 & ... & ...\\
    d$h6\eta_\mathrm{H}14$ & $6$ & $0$ & $14$ & 1428 & 923 & 0.65 & 21000 & ... & ...\\
    \noalign{\smallskip}
    \hline
    \end{tabular*}
    \tablefoot{Name, grid spacing $h$, Smagorinsky coefficient $C_\mathrm{S}$, artificial homogeneous magnetic diffusivity $\eta_\mathrm{H}$, effective Reynolds and magnetic Reynolds numbers $Re_\mathrm{eff}$ and $Re_{\mathrm{m},\mathrm{eff}}$, effective magnetic Prandtl number $Pr_{\mathrm{m},\mathrm{eff}}$, magnetic energy e-folding time $\tau_\mathrm{E}$, and time-averaged unsigned vertical magnetic field $\langle|B_z|\rangle$ and magnetic-to-kinetic energy ratio $\langle E_\mathrm{m}/E_\mathrm{k}\rangle$ (the last six parameters are computed as discussed in Sect.~\ref{sec:dynamo}).}
    \end{table*}
    
    The second set of simulations, with a seed field $B_z=1\ \mathrm{mG}$, is used in Sect.~\ref{sec:dynamo} to investigate dynamo action at $Pr_{\mathrm{m},\mathrm{eff}}\lesssim1$. To this aim, we carried out a total of ten simulations: five reference simulations with $h=24,16,12,8$, and $6\ \mathrm{km}$; three simulations with $h=6\ \mathrm{km}$, $C_\mathrm{S}=1.55$, and $\eta_\mathrm{H}=2.2,\ 2.5$, and $3.2\cdot10^{10}\ \mathrm{cm^2\,s^{-1}}$; and two simulations with $h=6\ \mathrm{km}$ and $\eta_\mathrm{H}=1.2$ and $1.4\cdot10^{10}\ \mathrm{cm^2\,s^{-1}}$. Because of some numerical instabilities emerging at low grid spacings, for all simulations with $h<16\ \mathrm{km}$ we introduced artificial numerical dissipation at the bottom boundary by setting $C_{\mathrm{lin}}=0.0025$ and $C_{\mathrm{sqrt}}=0.002$, which were further increased to $C_{\mathrm{lin}}=0.008$ and $C_{\mathrm{sqrt}}=0.004$ for the last 2 h of simulation d$h6$. An investigation of the impact of these two parameters on dynamo action is discussed in Sect.~\ref{sec:dynamo}. Table~\ref{tab:1mG} summarises the numerical setups of the dynamo runs.
    
\section{Evaluation of $\nu$ and $\eta$}\label{sec:res_etanu}
    In Sect.~\ref{sec:methodology}, it is explained that, after having obtained $\rho^h$, $\vec{v}^h$, and $\vec{B}^h$ (the numerical results from the simulations described in Sect.~\ref{sec:setup}), we need a post-processing scheme to discretise the operators $\delta_t^{\mathrm{pp},h},(\nabla\cdot)^{\mathrm{pp},h},(\nabla^2)^{\mathrm{pp},h}$, and $\nabla^{\mathrm{pp},h}$ and apply the methodology we propose. To ensure that the choice of the post-processing scheme does not influence our conclusions, in the following we consider three finite difference schemes of different order of accuracy: second, fourth, and sixth order, denoted as FD2, FD4, and FD6, respectively. Moreover, we need to choose how to express the term $\nabla\cdot\bm{\tau}$. To investigate if the particular form of $\nabla\cdot\bm{\tau}$ has an impact on the resulting effective viscosity, in the following we consider the two expressions:
    \begin{eqnarray}
       (\nabla\cdot\bm{\tau})^\mathrm{a}&=&\nu^\mathrm{a}\nabla\cdot\{\rho[\nabla\vec{v}+(\nabla\vec{v})^T-2\nabla\cdot\vec{v}/3]\}\label{eq:divtaua}\\ (\nabla\cdot\bm{\tau})^\mathrm{b}&=&\nu^\mathrm{b}\rho\nabla^2\vec{v}.\label{eq:divtaub}
    \end{eqnarray}
    The first expression corresponds to the typical form used to express the divergence of the stress tensor in the context of viscous flows, whereas the second expression represents a simple diffusion term in the velocity equation with diffusion coefficient $\nu^b$. We note that $\nu^a$ and $\nu^b$ are free parameters of the post-processing methodology that are used to estimate the effective viscosity affecting the numerical results, whereas $\nu_\mathrm{H}$ and $C_\mathrm{S}$ are input parameters of two versions of the stress tensor implemented in CO\textsuperscript{5}BOLD and used to add explicit viscosity to the momentum equation. 
    
    In Sect.~\ref{sec:methodology} it is assumed that part of the numerical errors introduced by the discretisation of the induction and momentum equations can be expressed as $\eta\nabla^2\vec{B}$ and $\nabla\cdot\bm{\tau}$, respectively. To investigate the validity of this assumption for the induction equation implemented in CO\textsuperscript{5}BOLD, we computed the Pearson correlation coefficient $r_B$ between $(\nabla^2)^{\mathrm{pp},h}\vec{B}^h$ and $\delta_t^{\mathrm{pp},h}\vec{B}^h-\sum_{i=1}^2w_iO_i^{\mathrm{pp},h}(\vec{v}^h,\vec{B}^h)$ (with weights and operators in Eq.~(\ref{eq:wOi})) for the simulations $h24,h16,h12,h8$, and $h6$ and at $t=500\ \mathrm{s}$ and $t=1000\ \mathrm{s}$.
    \begin{table*}\centering
    \caption{Pearson correlation coefficient $r_B$.}
    \label{tab:rB}
    \begin{tabular*}{\linewidth}{l@{\extracolsep{\fill}}cccccc}
    \hline\hline
    \noalign{\smallskip}
    \multirow{2}{*}{Simulation name} & \multicolumn{3}{c}{$t=500\ \mathrm{s}$} & \multicolumn{3}{c}{$t=1000\ \mathrm{s}$}\\
    \cmidrule{2-4}\cmidrule{5-7}
    & FD2 & FD4 & FD6 & FD2 & FD4 & FD6\\
    \noalign{\smallskip}
    \hline
    \noalign{\smallskip}
    $h24$ & 0.85 & 0.89 & 0.88 & 0.85 & 0.89 & 0.88\\
    $h16$ & 0.84 & 0.88 & 0.87 & 0.83 & 0.87 & 0.86\\
    $h12$ & 0.81 & 0.86 & 0.86 & 0.81 & 0.86 & 0.85\\
    $h8$  & 0.79 & 0.85 & 0.84 & 0.77 & 0.84 & 0.83\\
    $h6$  & 0.77 & 0.84 & 0.83 & 0.77 & 0.84 & 0.83\\
    \noalign{\smallskip}
    \hline
    \end{tabular*}
    \tablefoot{Simulation name and Pearson correlation coefficients $r_B$ at $t=500\ \mathrm{s}$ and $t=1000\ \mathrm{s}$ obtained using the post-processing schemes FD2, FD4, and FD6.}
    \end{table*}
    \begin{table*}\centering
    \caption{Pearson correlation coefficient $r_{\rho v}^a$.}
    \label{tab:rrhova}
    \begin{tabular*}{\linewidth}{l@{\extracolsep{\fill}}cccccc}
    \hline\hline
    \noalign{\smallskip}
    \multirow{2}{*}{Simulation name} & \multicolumn{3}{c}{$t=500\ \mathrm{s}$} & \multicolumn{3}{c}{$t=1000\ \mathrm{s}$}\\
    \cmidrule{2-4}\cmidrule{5-7}
    & FD2 & FD4 & FD6 & FD2 & FD4 & FD6\\
    \noalign{\smallskip}
    \hline
    \noalign{\smallskip}
    $h24$ & 0.65 & 0.66 & 0.66 & 0.67 & 0.69 & 0.68\\
    $h16$ & 0.63 & 0.65 & 0.64 & 0.62 & 0.64 & 0.63\\
    $h12$ & 0.59 & 0.61 & 0.61 & 0.61 & 0.63 & 0.63\\
    $h8$  & 0.58 & 0.60 & 0.59 & 0.58 & 0.60 & 0.60\\
    $h6$  & 0.57 & 0.61 & 0.60 & 0.57 & 0.59 & 0.59\\
    \noalign{\smallskip}
    \hline
    \end{tabular*}
    \tablefoot{Simulation name and Pearson correlation coefficients $r_{\rho v}^a$ at $t=500\ \mathrm{s}$ and $t=1000\ \mathrm{s}$ obtained using the post-processing schemes FD2, FD4, and FD6.}
    \end{table*}
    \begin{table*}\centering
    \caption{Pearson correlation coefficient $r_{\rho v}^b$.}
    \label{tab:rrhovb}
    \begin{tabular*}{\linewidth}{l@{\extracolsep{\fill}}cccccc}
    \hline\hline
    \noalign{\smallskip}
    \multirow{2}{*}{Simulation name} & \multicolumn{3}{c}{$t=500\ \mathrm{s}$} & \multicolumn{3}{c}{$t=1000\ \mathrm{s}$}\\
    \cmidrule{2-4}\cmidrule{5-7}
    & FD2 & FD4 & FD6 & FD2 & FD4 & FD6\\
    \noalign{\smallskip}
    \hline
    \noalign{\smallskip}
    $h24$ & 0.65 & 0.66 & 0.65 & 0.66 & 0.68 & 0.68\\
    $h16$ & 0.63 & 0.65 & 0.64 & 0.62 & 0.64 & 0.63\\
    $h12$ & 0.59 & 0.61 & 0.60 & 0.61 & 0.63 & 0.62\\
    $h8$  & 0.57 & 0.60 & 0.59 & 0.57 & 0.60 & 0.59\\
    $h6$  & 0.57 & 0.60 & 0.60 & 0.56 & 0.59 & 0.58\\
    \noalign{\smallskip}
    \hline
    \end{tabular*}
    \tablefoot{Simulation name and Pearson correlation coefficients $r_{\rho v}^b$ at $t=500\ \mathrm{s}$ and $t=1000\ \mathrm{s}$ obtained using the post-processing schemes FD2, FD4, and FD6.}
    \end{table*}
    The results are displayed in Table~\ref{tab:rB}. We find that $r_B>0.76$ independently of the order of accuracy of the post-processing scheme FD2-FD6 used for the computation. We repeated the same analysis for the momentum equation, for which we computed the Pearson correlation coefficients $r_{\rho v}^a$ and $r_{\rho v}^b$ between $\nabla\cdot\{\rho[\nabla\vec{v}+(\nabla\vec{v})^T-2\nabla\cdot\vec{v}/3]\}$ and $\delta_t^{\mathrm{pp},h}(\rho\vec{v}^h)-\sum_{i=1}^4w_iO_i^{\mathrm{pp},h}(\rho^h,\vec{v}^h,\vec{B}^h)$, and between $\rho\nabla^2\vec{v}$ and $\delta_t^{\mathrm{pp},h}(\rho\vec{v}^h)-\sum_{i=1}^4w_iO_i^{\mathrm{pp},h}(\rho^h,\vec{v}^h,\vec{B}^h)$, respectively (with weights and operators in Eq.~(\ref{eq:wOm})), again for the simulations $h24,h16,h12,h8$, and $h6$ and at $t=500\ \mathrm{s}$ and $t=1000\ \mathrm{s}$. The results are displayed in Tables~\ref{tab:rrhova} and \ref{tab:rrhovb}. It is found that $r_{\rho v}^a,r_{\rho v}^b>0.56$. Although the Pearson correlation coefficients for the momentum equation are smaller than for the induction equation, $r_{\rho\varv}^a$, $r_{\rho\varv}^b$, and $r_B$ are noticeably larger than 0, implying that a considerable linear correlation exists between the dissipation terms and the residuals of the corresponding equations. This is consistent with our assumption that the residuals can be, to a large degree, expressed in the form of diffusion terms.

    We can now apply the methodology of Sect.~\ref{sec:methodology} and estimate $\nu_\mathrm{eff}$ and $\eta_\mathrm{eff}$. To this aim, in the following, we minimise the residual with the procedures (i) and (ii), Eqs.~(\ref{eq:i}) and (\ref{eq:ii}), considering all the spatial grid points on a horizontal plane at once and plane by plane independently, thus obtaining effective dissipation coefficients that depend on time and height, $\nu_\mathrm{eff}(t,z)$ and $\eta_\mathrm{eff}(t,z)$.
    \begin{figure*}
    \centering
    \includegraphics[width=17cm]{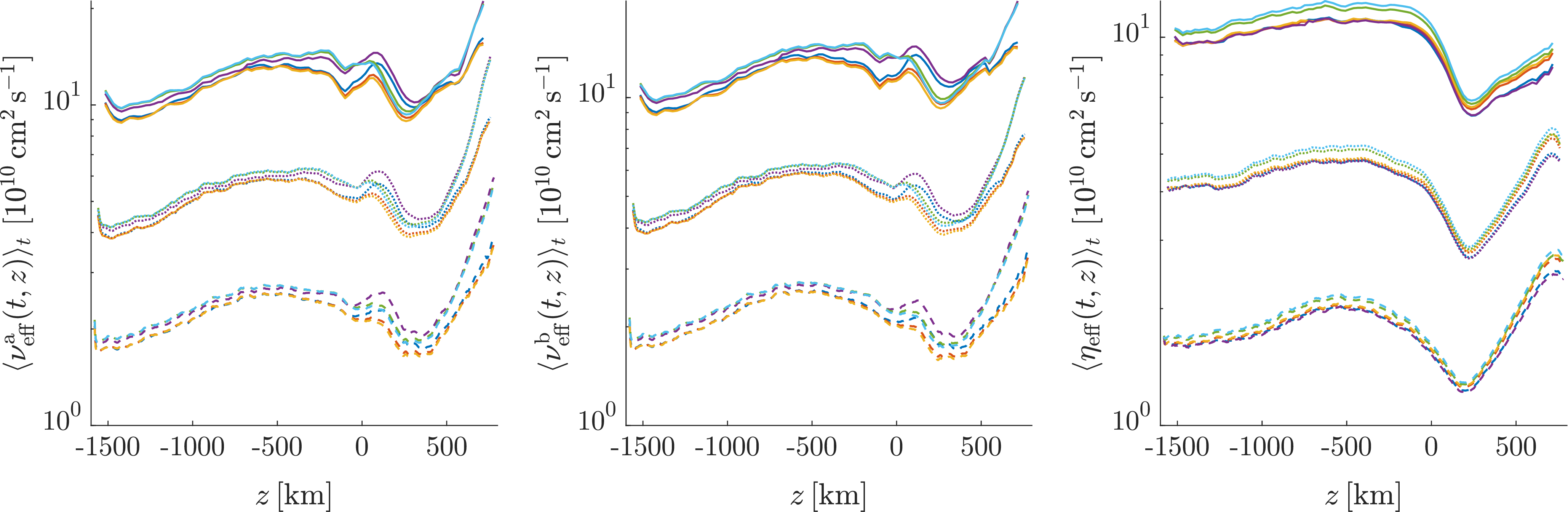}
      \caption{Effective viscosities and plasma resistivity as a function of height. Time-averaged profiles of $\nu^\mathrm{a}_\mathrm{eff}(t,z)$ (\emph{left panel}), $\nu^\mathrm{b}_\mathrm{eff}(t,z)$ (\emph{centre panel}), and $\eta_\mathrm{eff}(t,z)$ (\emph{right panel}) for the simulations $h24$ (solid lines), $h12$ (dotted lines), and $h6$ (dashed lines). Colour code: blue, red, and yellow for method (i) with post-processing scheme FD2, FD4, and FD6, respectively, and purple, green, and light blue for method (ii) with post-processing scheme FD2, FD4, and FD6, respectively.
              }
         \label{Fig:coeffs_z}
    \end{figure*}
    In Fig.~\ref{Fig:coeffs_z} we display $\langle\nu_\mathrm{eff}(t,z)\rangle_t$ and $\langle\eta(t,z)_\mathrm{eff}\rangle_t$, where the time averages are performed over the time instants $t=500\ \mathrm{s}$ and $t=1000\ \mathrm{s}$, computed with both procedures (i) and (ii) and considering both forms of $\nabla\cdot\bm{\tau}$ in Eqs.~(\ref{eq:divtaua}) and (\ref{eq:divtaub}), for the simulations $h24,h12$, and $h6$. As we could expect, the effective viscosity and magnetic diffusivity decrease by refining the grid. Moreover, it appears that the profiles vary less in the convection zone ($z<0\ \mathrm{km}$) than in the photosphere ($z>0\ \mathrm{km}$), the main exception being $\nu_\mathrm{eff}$ near the bottom boundary. We think that the steep increase 
    in the upper photosphere, above $z=300\ \mathrm{km}$, is due to the presence of shock waves in this region, which lead to larger gradients of the plasma velocity and of the magnetic field, and therefore to larger dissipation coefficients.\footnote{In the vicinity of shock fronts, a shock capturing numerical scheme automatically reduces the order of accuracy to first order,  which increases the effective diffusivity of the scheme.} We also see that the results do not have a strong dependence on the post-processing scheme employed to compute the coefficients, except for some small local deviations and for the evaluation of $\eta_\mathrm{eff}$ with the procedure (ii). Also, $\nu_\mathrm{eff}$ is almost independent of the form chosen to express $\nabla\cdot\bm{\tau}$, that is, the left and middle panels of Fig.~\ref{Fig:coeffs_z} are almost identical. Concerning the differences resulting from employing Eq.~(\ref{eq:i}) or Eq.~(\ref{eq:ii}), procedures (i) or (ii), we observe that the second approach generally produces slightly larger coefficients, as we could expect according to our remark in Sect.~\ref{sec:methodology}. Nevertheless, the differences are rather small, in particular in the convection zone where they are typically smaller than $15\%$. We note that very similar trends are obtained for the simulations $h16$ and $h8$ (not displayed in Fig.~\ref{Fig:coeffs_z} for clarity). We also tested a different approach for minimising Eq.~(\ref{eq:i}) or Eq.~(\ref{eq:ii}): we considered all the spatial grid points in the numerical domain at once, instead of independently plane by plane, thus obtaining global dissipation coefficients that only depend on time (not shown here for conciseness). They confirm the trends discussed above, with marginal dependence on the post-processing scheme and resulting coefficients that are slightly larger for procedure (ii) than for (i), with differences that are generally smaller than $10\%$.
    
    Given these results, in the following we consider only the post-processing scheme FD2, the methodology (ii), and we express $\nabla\cdot\bm{\tau}$ according to Eq.~(\ref{eq:divtaub}) (we omit the superscript $\mathrm{b}$ to lighten the notation). We note that, in order to solve Eq.~(\ref{eq:ii}) for $\nu_\mathrm{eff}(t,z)$ and $\eta_\mathrm{eff}(t,z)$, we implemented the necessary routines directly in CO\textsuperscript{5}BOLD, such that we can compute the dissipation coefficients at run time with much higher output frequency, without the need for storing the full snapshots at these high output frequencies. In the remainder of this section, we consider an output frequency of the numerical results of $5\ \mathrm{s}$, with time averages over the interval $t\in[500,1000]\ \mathrm{s}$. The standard deviations of the time signals of the dissipation coefficients $\nu_\mathrm{eff}(t,z)$ and $\eta_\mathrm{eff}(t,z)$ during this interval are rather small. They are generally below $5\%$, except at certain locations, in particular near the top and bottom boundaries, where they can increase to 10-20\%.

    \begin{figure*}
    \centering
    \includegraphics[width=17cm]{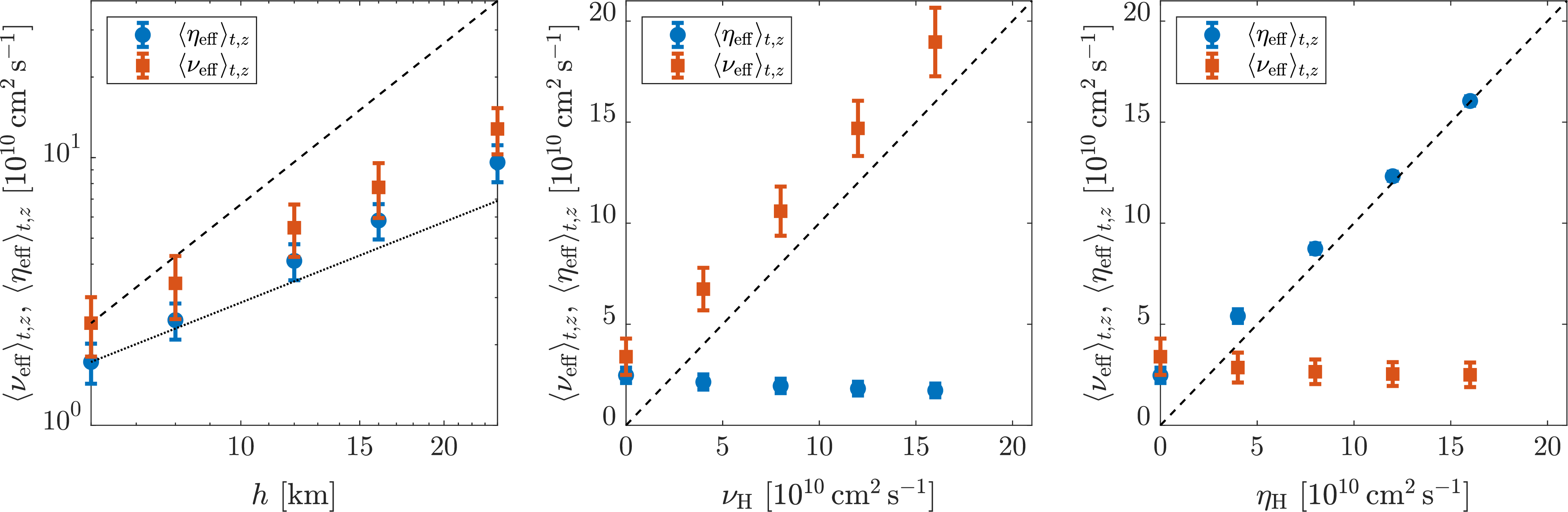}
      \caption{Time and spatial averages of $\nu_\mathrm{eff}$ (red squares) and $\eta_\mathrm{eff}$ (blue circles) for the simulations $h24,h16,h12,h8$, and $h6$ as function of grid spacing $h$ (\emph{left panel}); for the simulations $h8$, $h8\nu_\mathrm{H}4$, $h8\nu_\mathrm{H}8$, $h8\nu_\mathrm{H}12$, and $h8\nu_\mathrm{H}16$ as function of $\nu_\mathrm{H}$ (\emph{centre panel}); and for the simulations $h8$, $h8\eta_\mathrm{H}4$, $h8\eta_\mathrm{H}8$, $h8\eta_\mathrm{H}12$, and $h8\eta_\mathrm{H}16$ as function of $\eta_\mathrm{H}$ (\emph{right panel}). The dashed and dotted black lines in the \emph{left panel} are proportional to $h^2$ and $h$, respectively, whereas the dashed black lines in the \emph{centre and right panels} are proportional to $\nu_\mathrm{H}$ and $\eta_\mathrm{H}$, respectively. The error bars denote one standard deviation computed over time and vertical direction $z$.
              }
         \label{Fig:coeffs_hnueta}
    \end{figure*}

    The PoPe methodology states that if a code is bug free, then $\delta w_i$ should vanish at rate $p$ in the asymptotic limit $h\rightarrow0$.
    This should also hold true for $\nu_\mathrm{eff}$ and $\eta_\mathrm{eff}$. To verify if it is the case for CO\textsuperscript{5}BOLD, Fig.~\ref{Fig:coeffs_hnueta} (left) displays the effective viscosity and magnetic diffusivity for the simulations $h24,h16,h12,h8$, and $h6$, averaged over the vertical coordinate $z$, as a function of the grid spacing $h$. The dissipation coefficients are clearly decreasing as we refine the grid, at a rate between 1 (linear) and 2 (quadratic). This is consistent with the order of accuracy of the FRweno scheme implemented in CO\textsuperscript{5}BOLD. We remind the reader that this result is independent of the order of accuracy of the post-processing scheme used to compute $\nu_\mathrm{eff}$ and $\eta_\mathrm{eff}$; thus, it indicates that we are not measuring an extra dissipation introduced by our methodology. We repeated the same analysis for the simulations $h8\nu_\mathrm{H}4$, $h8\nu_\mathrm{H}8$, $h8\nu_\mathrm{H}12$, and $h8\nu_\mathrm{H}16$ (Fig.~\ref{Fig:coeffs_hnueta}, centre), and for the simulations $h8\eta_\mathrm{H}4$, $h8\eta_\mathrm{H}8$, $h8\eta_\mathrm{H}12$, and, $h8\eta_\mathrm{H}16$ (Fig.~\ref{Fig:coeffs_hnueta}, right), displaying the results as function of $\nu_\mathrm{H}$ and $\eta_\mathrm{H}$, respectively. We obtain $\eta_\mathrm{eff}\simeq\eta_\mathrm{H}$ for large enough artificial magnetic diffusivities, whereas $\nu_\mathrm{eff}$ and $\eta_\mathrm{eff}$ converge to the intrinsic numerical diffusivity of the scheme for $\nu_\mathrm{H},\eta_\mathrm{H}\rightarrow0$. The effective numerical viscosity $\nu_\mathrm{eff}$ is slightly larger than $\nu_\mathrm{H}$ for finite artificial viscosities. In Fig.~\ref{Fig:coeffs_hnueta}, we also observe that $\eta_\mathrm{eff}$ decreases with increasing $\nu_\mathrm{H}$, and similarly $\nu_\mathrm{eff}$ decreases with increasing $\eta_\mathrm{H}$. We speculate that this is related to the frozen-in condition exhibited by plasmas under almost ideal conditions, which entails the smooth behaviour of the magnetic field with increasing viscosity and the smooth behaviour of the flow with increasing magnetic diffusivity.
    
    To gain a deeper insight into the methodology of Sect.~\ref{sec:methodology}, it is useful to remember that, according to the classical picture of turbulent energy cascade from Kolmogorov's theory, the kinetic energy that resides mostly at large scales is transferred across scales by non-linear processes and is eventually dissipated in the form of heat at about the Kolmogorov dissipation scale, $\lambda=(\nu^3/\varepsilon)^{1/4}$, where $\varepsilon$ is the rate of kinetic energy dissipation per mass unit~\citep{Kolmogorov1941}. Assuming $\varepsilon(t,z)=2\nu_\mathrm{eff}(t,z)\int_0^\infty E_\mathrm{v}(k_\mathrm{h},t,z)k_\mathrm{h}^2dk_\mathrm{h}$, with $E_\mathrm{v}(k_\mathrm{h},t,z)=\hat{\vec{v}}(k_\mathrm{h},t,z)\cdot\hat{\vec{v}}^*(k_\mathrm{h},t,z)/2$ the velocity spectrum, $k_\mathrm{h}$ the horizontal wave number, and $\hat{\vec{v}}$ and $\hat{\vec{v}}^*$ the Fourier transform of $\vec{v}$ and its complex conjugate, respectively, we can compute the Kolmogorov dissipation scale for the simulations $h24,h16,h12,h8$, and $h6$. We obtain $\langle\lambda\rangle_{t,z}=23.6,16.5,12.6,8.8$, and $6.9\ \mathrm{km}$, that is, $\langle\lambda\rangle_{t,z}\approx h$. This is consistent with an intrinsic viscosity acting approximately at the grid scale.

    \begin{figure*}
    \centering
    \includegraphics[width=17cm]{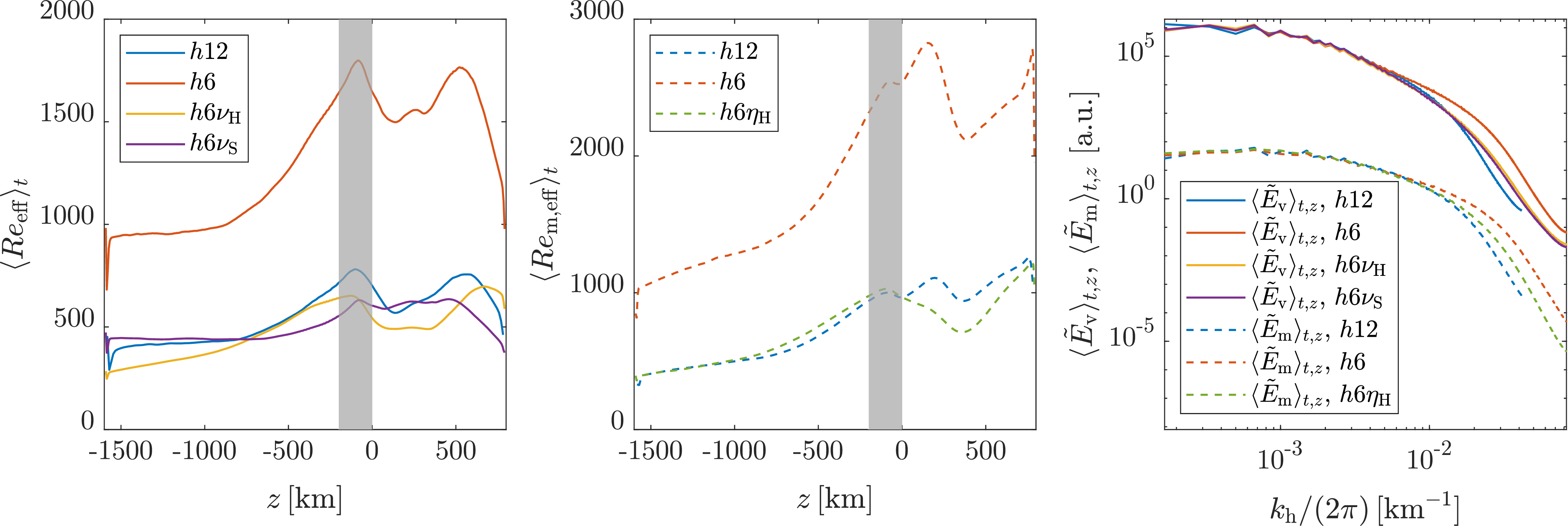}
      \caption{Reynolds and magnetic Reynolds numbers as a function of height and kinetic and magnetic energy spectra. \emph{Left:} Time-averaged profiles of $Re_\mathrm{eff}(t,z)$ as function of geometrical height $z$ for the simulations $h12$, $h6$, $h6\nu_\mathrm{H}$, and $h6\nu_\mathrm{S}$. \emph{Centre:} Time-averaged profiles of $Re_{\mathrm{m},\mathrm{eff}}(t,z)$ for the simulations $h12$, $h6$, and $h6\eta_\mathrm{H}$. \emph{Right:} Spectra of $\tilde{E}_\mathrm{v}$ (solid lines) and $\tilde{E}_\mathrm{m}$ (dashed lines) averaged over time and in space between $z=-200\ \mathrm{km}$ and $z=0\ \mathrm{km}$ as a function of horizontal wave number $k_\mathrm{h}$. The spectra of $\tilde{E}_\mathrm{v}$ are multiplied by $10^4$ for clarity. The shaded grey areas of the \emph{left and centre panels} denote the region between $z=-200\ \mathrm{km}$ and $z=0\ \mathrm{km}$, used for the spatial average of the energy spectra.
              }
         \label{Fig:Re_Ek}
    \end{figure*}

    \begin{figure}
    \centering
    \includegraphics[width=\linewidth]{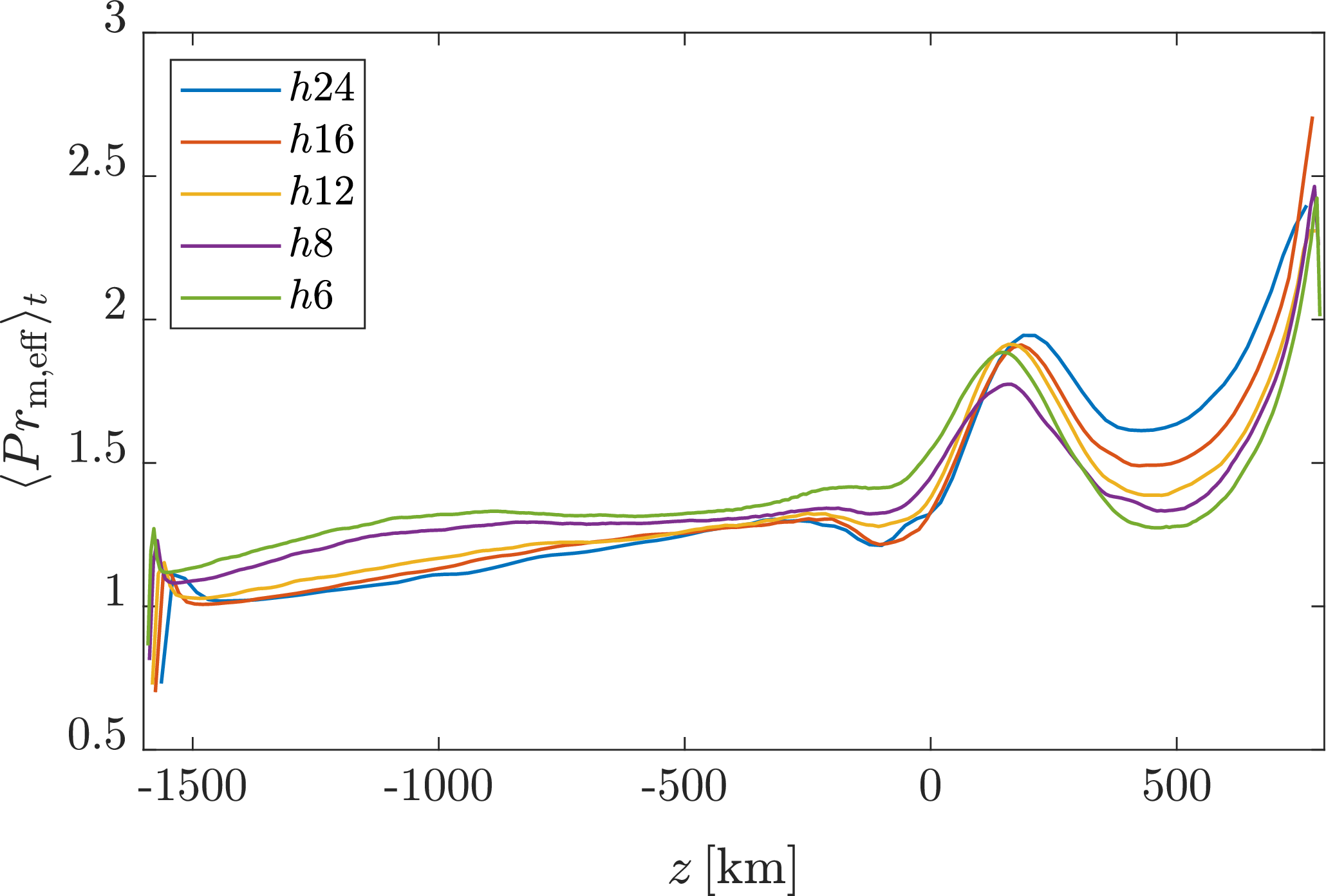}
      \caption{Time-averaged profiles of the effective magnetic Prandtl number $Pr_{\mathrm{m},\mathrm{eff}}$ as a function of geometrical height $z$ for the simulations $h24$, $h16$, $h12$, $h8$, and $h6$.
              }
         \label{Fig:Prm}
    \end{figure}

    In the context of Kolmogorov's theory it is also often assumed that, for large enough $Re$ and $Re_\mathrm{m}$, the statistics of velocity fluctuations at small scales is uniquely determined by $\nu$ and $\varepsilon$, and that $\eta$ affects the distribution of magnetic fields only at the smallest scales~\citep{Thaler2015}.
    Accordingly, two simulations with similar $Re_\mathrm{eff}(t,z)=L\varv_\mathrm{rms}(t,z)/\nu_\mathrm{eff}(t,z)$, where $L=1\ \mathrm{Mm}$ is the typical length scale and $\varv_\mathrm{rms}(t,z)=\sqrt{\langle|\vec{v}|^2\rangle_\mathrm{h}}$ is the root-mean-square value of the fluid velocity, should exhibit similar velocity spectra; analogously, two simulations with similar $Re_{\mathrm{m},\mathrm{eff}}(t,z)=L\varv_\mathrm{rms}(t,z)/\eta_\mathrm{eff}(t,z)$ should have similar magnetic energy spectra. Consequently, a high-resolution simulation with explicit viscosity and magnetic diffusivity is expected to behave like a lower resolution simulation of similar $Re_\mathrm{eff}$ and $Re_{\mathrm{m},\mathrm{eff}}$.
    To investigate this hypothesis, we carried out the simulations $h6\nu_\mathrm{H}$ and $h6\nu_\mathrm{S}$, adjusting $\nu_\mathrm{H}$ and $C_\mathrm{S}$ to approximately replicate the effective viscosity of the simulation $h12$ (similarly, we adjusted $\eta_\mathrm{H}$ in $h6\eta_\mathrm{H}$ to replicate the effective magnetic diffusivity of simulation $h12$). The resulting vertical profiles of $\langle Re_\mathrm{eff}\rangle_t$ and $\langle Re_{\mathrm{m},\mathrm{eff}}\rangle_t$ are displayed in Fig.~\ref{Fig:Re_Ek}, left and centre, respectively. While the profiles exhibit some local differences, which can be expected due to the different nature of the dissipation terms introduced in the different simulations, the overall agreement is rather good, in particular if compared to the results of simulation $h6$, displayed in red in the aforementioned panels of Fig.~\ref{Fig:Re_Ek}. Figure~\ref{Fig:Re_Ek} (right) displays the velocity and magnetic energy spectra, normalised to their integral area, $\tilde{E}_\mathrm{v}$ and $\tilde{E}_\mathrm{m}$, averaged over time and between $z=-200\ \mathrm{km}$ and $z=0\ \mathrm{km}$. Concerning the velocity spectra, there is indeed good agreement among the simulations $h12$, $h6\nu_\mathrm{H}$ and $h6\nu_\mathrm{S}$, although some local differences are present. These spectra are clearly offset from the velocity spectrum of simulation $h6$. Concerning the magnetic spectra, there is a better agreement between $h6\eta_\mathrm{H}$ and $h12$ than between $h6$ and $h12$, although noticeable differences appear at small scales. These results reveal that it is not enough to replicate the values of $Re_\mathrm{eff}$ and $Re_{\mathrm{m},\mathrm{eff}}$, or equivalently $\nu_\mathrm{eff}$ and $\eta_\mathrm{eff}$, to obtain the same spectra; the nature of the dissipation terms plays a role in setting the profiles of $\tilde{E}_\mathrm{v}$ and $\tilde{E}_\mathrm{m}$. Evaluating the average velocity and magnetic spectra in different depths in the convection zone (not shown here) yielded the same results.

    In light of the results discussed in the present section, we are now confident that our methodology provides a reliable estimate of $\nu_\mathrm{eff}$ and $\eta_\mathrm{eff}$.
    We can therefore compute the magnetic Prandtl number, $Pr_{\mathrm{m},\mathrm{eff}}=\nu_\mathrm{eff}/\eta_\mathrm{eff}$, as discussed in Sect.~\ref{sec:methodology}. The time-averaged results for the simulations $h24$, $h16$, $h12$, $h8$, and $h6$ are displayed in Fig.~\ref{Fig:Prm}. The profiles of $\langle Pr_{\mathrm{m},\mathrm{eff}}\rangle_t$ are quite flat in the convection zone, with a value slightly above one, the only exception being near the bottom boundary, where the magnetic Prandtl number drops below one. There is a trend related to the grid resolution, with $\langle Pr_{\mathrm{m},\mathrm{eff}}\rangle_t$ slightly increasing in the convection zone with finer grids, whereas the opposite occurs in the photosphere.

\section{Small-scale dynamo simulations}\label{sec:dynamo}
    Dynamo runs with increasing spatial resolution, and consequently increasing Reynolds and magnetic Reynolds numbers, were carried out in order to investigate the possibility of simulating small-scale dynamo action with CO\textsuperscript{5}BOLD. The corresponding runs and numerical parameters are summarised in Table~\ref{tab:1mG}. There, the Reynolds and magnetic Reynolds numbers, as well as the magnetic Prandtl numbers, are estimated as the time averages over the kinematic phase and the spatial averages for $z<0\ \mathrm{km}$ of $Re_\mathrm{eff}(z,t)$, $Re_{\mathrm{m},\mathrm{eff}}(z,t)$, and $Pr_{\mathrm{m},\mathrm{eff}}(z,t)=Re_{\mathrm{m},\mathrm{eff}}(z,t)/Re_\mathrm{eff}(z,t)$, respectively.

    \begin{figure*}
    \centering
    \includegraphics[width=17cm]{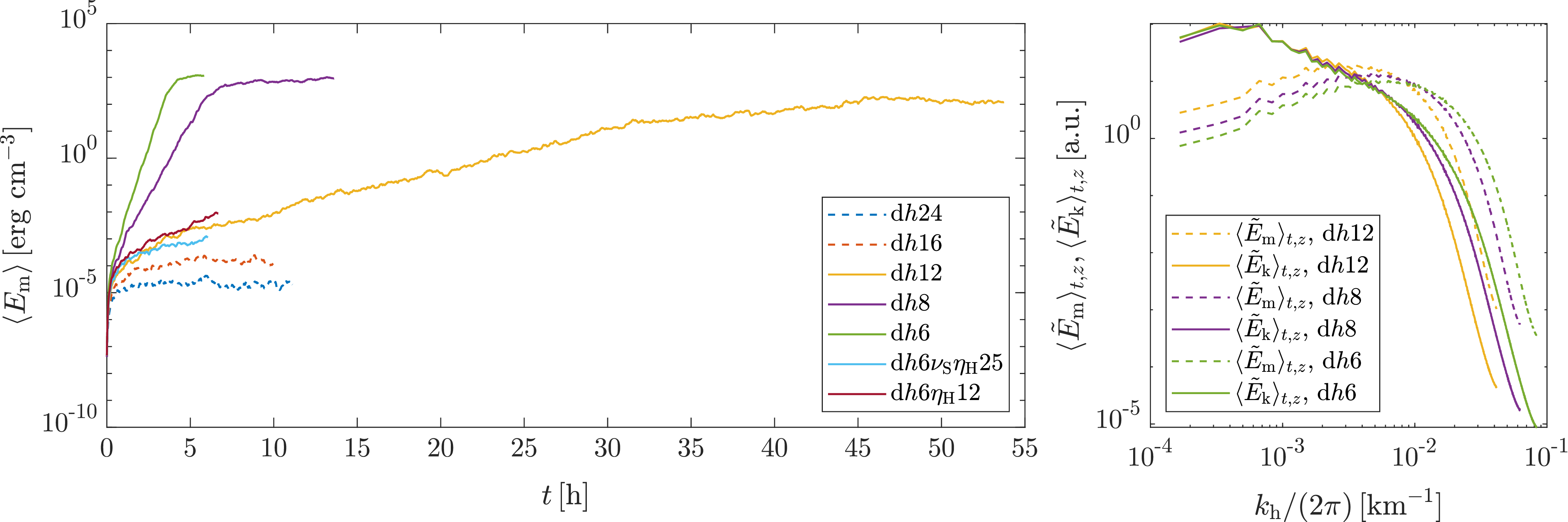}
      \caption{Growth of magnetic energy with time and kinetic and magnetic spectra. \emph{Left:} Temporal profiles of the magnetic energy density averaged over the full domain as function of time for simulations $\mathrm{d}h24$, $\mathrm{d}h16$, $\mathrm{d}h12$, $\mathrm{d}h8$, $\mathrm{d}h6$, $\mathrm{d}h6\nu_\mathrm{s}\eta_\mathrm{H}25$, and $\mathrm{d}h6\eta_\mathrm{H}12$. Solid lines denote simulations showing dynamo action, whereas dashed lines denote simulations without dynamo action. \emph{Right:} Spectra of the normalised kinetic energy density $\tilde{E}_\mathrm{k}$ (solid lines) and magnetic energy density $\tilde{E}_\mathrm{m}$ (dashed lines) averaged over the time span of the kinematic phase and in space between $z=-900\ \mathrm{km}$ and $z=-700\ \mathrm{km}$.
              }
         \label{Fig:Em_t}
    \end{figure*}

    The resulting time evolution of the magnetic energy density, averaged over the full domain, for the five simulations $\mathrm{d}h24$, $\mathrm{d}h16$, $\mathrm{d}h12$, $\mathrm{d}h8$, and $\mathrm{d}h6$, is displayed in Fig.~\ref{Fig:Em_t}, left.
    The simulations start with a short transient phase of a few minutes where magnetic flux is expelled from the granules and transported to the intergranular space. After that, the temporal profiles of $\langle E_\mathrm{m}\rangle$ are rather flat for $h=16\ \mathrm{km}$ and $h=24\ \mathrm{km}$ (dashed lines), and the overall magnetic energy density remains extremely low. On the other hand, for grids with $h<16\ \mathrm{km}$, we clearly recognise a linear (kinematic) phase, during which magnetic energy is amplified exponentially, followed by a non-linear (saturation) phase occurring once the impact of the Lorentz force on the fluid motion starts to be significant. The magnetic energy e-folding times reported in Table~\ref{tab:1mG}, $\tau_E$, are then obtained by fitting the $\langle E_m\rangle$ temporal profiles during the kinematic phase with an exponential function. As one would expect for a small-scale dynamo, the growth rate, $\gamma_E=1/\tau_E$, is an increasing function of $Re_{\mathrm{m},\mathrm{eff}}$, with $\gamma_E\sim Re_{\mathrm{m},\mathrm{eff}}$, which is consistent with what was found in~\citet{Graham2010} and \citet{Rempel2014}. A detailed investigation of the time evolution of the magnetic energy density in planes of different heights in the convection zone (not shown here for conciseness) yields energy e-folding times very similar to those presented in Table~\ref{tab:1mG}, which refer to the full domain. On the other hand, the magnetic energy is amplified much slower in the upper photosphere. For this reason, in the following we focus our attention to the convection zone alone, where we expect a more effective dynamo.
    
    Figure~\ref{Fig:Em_t} (right) shows the kinetic and magnetic energy spectra $\langle\tilde{E}_\mathrm{k}\rangle_{t,z}$ and $\langle\tilde{E}_\mathrm{m}\rangle_{t,z}$, normalised to their integral area and averaged over time during the kinematic phase and in space between $z=-900\ \mathrm{km}$ and $z=-700\ \mathrm{km}$, for the simulations $\mathrm{d}h12$, $\mathrm{d}h8$, and $\mathrm{d}h6$. Clearly, the magnetic energy spectra peak at scales much smaller than the kinetic energy spectra, which peak at a scale of about $1.5\ \mathrm{Mm}$, and their peaks shift to smaller scales when increasing the resolution. This is consistent with small-scale dynamo action. However, without further analysis it cannot be excluded that the observed amplification of small-scale magnetic fields originates from mechanisms other than the action of a small-scale dynamo, such as a turbulent tangling (or shredding) of the large-scale magnetic fields that could be produced by a mean-flow ($1\ \mathrm{Mm}$ scale) dynamo or by the Alfvénic response of large-scale fields to small-scale velocity fluctuations (turbulent induction). To rule out these possibilities and disentangle the different sources of small-scale magnetic energy, we follow the approach proposed in~\citet{Graham2010} and \citet{Rempel2014} and carry out a spectral analysis of the energy exchanges between kinetic and magnetic energy reservoirs. More precisely, as explained in the appendices of~\citet{Graham2010} and \citet{Rempel2014}, starting from the induction equation, we can decompose the time evolution of the spectral magnetic energy, $E_\mathrm{m}(k_\mathrm{h})=\hat{\vec{B}}(k_\mathrm{h})\cdot\hat{\vec{B}}^*(k_\mathrm{h})/(8\pi)$, in three contributions converting kinetic energy into magnetic energy and vice versa as 
    \begin{equation}\label{eq:dtEm}
        \partial_tE_\mathrm{m}(k_\mathrm{h})\approx T_\mathrm{ms}(k_\mathrm{h})+T_\mathrm{ma}(k_\mathrm{h})+T_\mathrm{md}(k_\mathrm{h}),
    \end{equation}
    where 
    \begin{align}
        T_\mathrm{ms}(k_\mathrm{h})&=\hat{\vec{B}}(k_\mathrm{h})\cdot\widehat{\nabla\cdot(\vec{Bv})}^*(k_\mathrm{h})/(8\pi)+\mathrm{c.c.,}\\
        T_\mathrm{ma}(k_\mathrm{h})&=-\hat{\vec{B}}(k_\mathrm{h})\cdot\widehat{\nabla\cdot(\vec{vB})}^*(k_\mathrm{h})/(8\pi)+\mathrm{c.c., and}\\
        T_\mathrm{md}(k_\mathrm{h})&=\eta_\mathrm{eff}\hat{\vec{B}}(k_\mathrm{h})\cdot\widehat{\nabla^2\vec{B}}^*(k_\mathrm{h})/(8\pi)+\mathrm{c.c.}
    \end{align}
    are the energy transfers to and from the magnetic energy due to the misalignment between velocity shear and magnetic field lines (stretching), to advection and compression, and to magnetic diffusivity, respectively, with $\hat{\text{-}}$ denoting the Fourier transform, $^*$ the complex conjugate, and $\mathrm{c.c.}$ the respective complex conjugate expression. The approximation symbol in Eq.~(\ref{eq:dtEm}) highlights the fact that this is not exactly equal, because $T_\mathrm{md}$ is an estimate of the magnetic energy losses due to numerical dissipation obtained under the assumption that the residual of the induction equation can be modelled with a Laplacian magnetic diffusivity. In the following, all the terms related to the energy transfer rates are computed using second-order, centred, finite difference schemes.
    \begin{figure*}
    \centering
    \includegraphics[width=17cm]{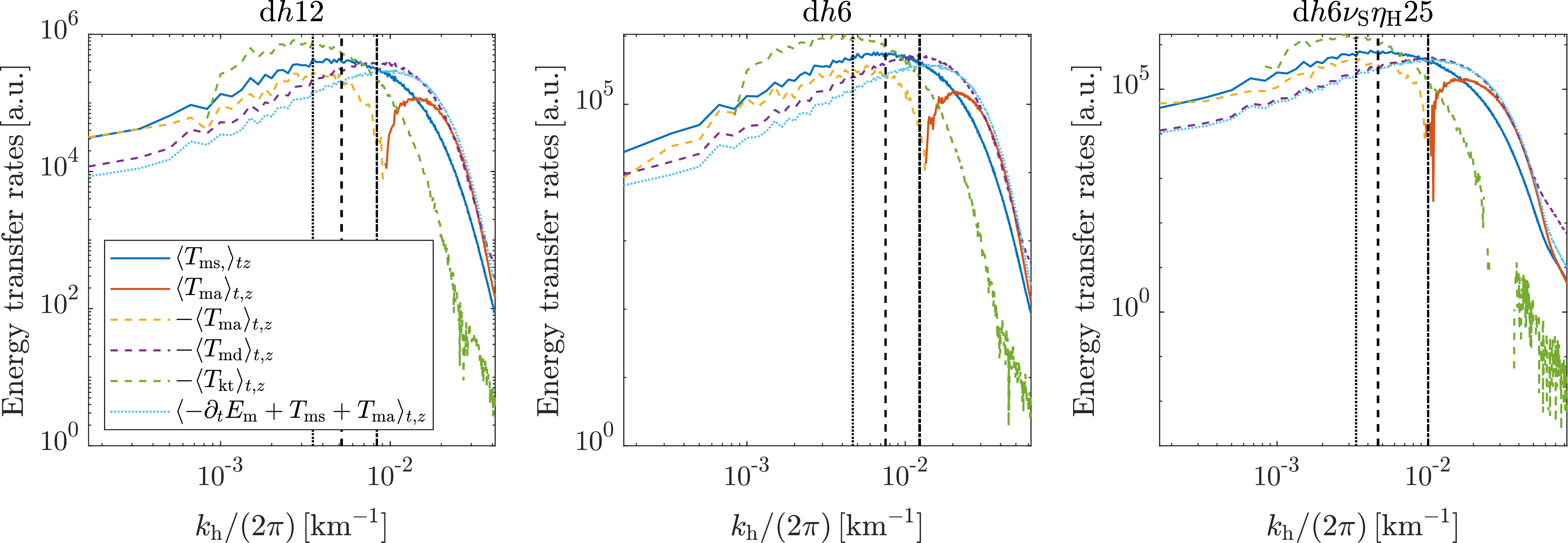}
      \caption{Normalised energy transfer functions averaged over the time span of the kinematic phase and in space between $z=-900\ \mathrm{km}$ and $z=-700\ \mathrm{km}$. \emph{Left:} for the simulation $\mathrm{d}h12$; \emph{centre:} for the simulation $\mathrm{d}h6$; and \emph{right:} for the simulation $\mathrm{d}h6\nu_\mathrm{S}\eta_\mathrm{H}25$. Solid colour curves represent sources of magnetic energy due to stretching (blue curves) and advection and compression (red curves) of magnetic fields, dashed colour curves represent magnetic energy losses due to advection and compression (yellow curves) and a Laplacian diffusion (purple curves) of magnetic fields. The green dashed curves represent the kinetic energy losses due to stretching of magnetic fields, whereas the light blue dotted lines represent the energy transfer imbalance $-\partial_tE_\mathrm{m}+T_\mathrm{ms}+T_\mathrm{ma}$. Black vertical lines denote the position of $\lambda_\mathrm{ms}$ (dashed), $\lambda_\mathrm{kt}$ (dotted), and $\lambda_\mathrm{md}$ (dash-dotted).
              }
         \label{Fig:TF}
    \end{figure*}
    
    Figure~\ref{Fig:TF}, left and centre, displays the spectral energy transfer rates, normalised to the magnetic energy density and averaged over time during the kinematic phase and in space between $z=-900\ \mathrm{km}$ and $z=-700\ \mathrm{km}$, for the two simulations $\mathrm{d}h12$ and $\mathrm{d}h6$, respectively. From $T_\mathrm{ms}$ (blue solid curve), we see that sources of magnetic energy due to the stretching of magnetic field lines are present at all scales of the simulations and that $T_\mathrm{ms}$ peaks on scales, $\lambda=2\pi/k_\mathrm{h}$, of about $\lambda_\mathrm{ms}\simeq200\ \mathrm{km}$ and $\lambda_\mathrm{ms}\simeq140\ \mathrm{km}$ for simulations $\mathrm{d}h12$ and $\mathrm{d}h6$, respectively; that is, the dominant scale for magnetic energy production by stretching is approximately $\lambda_\mathrm{ms}\simeq17\,h$ to $\lambda_\mathrm{ms}\simeq23\,h$ (black vertical dashed lines in Fig.~\ref{Fig:TF}). On the other hand, energy transfer by advection and compression results in a removal of magnetic energy from large scales (yellow dashed curves), breaking it down into smaller scale magnetic fields (solid red curves). At scales larger than $28\,h$, the term $T_\mathrm{ms}$ is mostly compensated by $T_\mathrm{ma}$ (solid blue versus dashed yellow curves). The contributions from numerical dissipation, shown as dashed purple curves in Fig.~\ref{Fig:TF}, dominate at small scales, approximately at $\lambda_\mathrm{md}\simeq10-14\,h$ (black vertical dash-dotted lines in Fig.~\ref{Fig:TF}). Additionally, from the momentum equation, we compute the energy transfer term representing kinetic energy losses due to the work of the Lorenz force via magnetic tension,
    \begin{equation}
    \begin{split}
        -T_\mathrm{kt}(k_\mathrm{h})=&-[\hat{\vec{v}}(k_\mathrm{h})\cdot\widehat{\nabla\cdot(\vec{BB})}^*(k_\mathrm{h})]/(16\pi)\\
        &-[\widehat{\rho\vec{v}}^*(k_\mathrm{h})\cdot\widehat{\nabla\cdot(\vec{BB})/\rho}(k_\mathrm{h})]/(16\pi)+\mathrm{c.c..}
    \end{split}
    \end{equation}
    This is displayed in Fig.~\ref{Fig:TF} in green. From $T_\mathrm{kt}$, we see that the stretching of magnetic field lines is responsible for losses of kinetic energy, transferred to the magnetic energy reservoir, at scales of approximately $\lambda_\mathrm{kt}\simeq24\,h$ to $\lambda_\mathrm{kt}\simeq36\,h$ (black vertical dotted lines in Fig.~\ref{Fig:TF}).
    
    From this analysis, it is clear that a range of scales exists between $\lambda_\mathrm{ms}$ (the peak of $T_\mathrm{ms}$) and $\lambda_\mathrm{kt}$ (the peak of $T_\mathrm{kt}$), where magnetic energy injection is mostly due to turbulent stretching of magnetic field lines, with the injected energy coming from the kinetic energy reservoir. Moreover, the magnetic energy spectra in Fig.~\ref{Fig:Em_t}, right, peak on scales of approximately $\lambda_{E_\mathrm{m}}\simeq27\,h$. Since these three scales, $\lambda_\mathrm{ms}$, $\lambda_{E_\mathrm{m}}$, and $\lambda_\mathrm{kt}$, all lie in the inertial range (see Fig.~\ref{Fig:Em_t}, right) and are incompatible with small-scale velocity fluctuations interacting with a large-scale magnetic field to produce small-scale magnetic energy, we conclude that the magnetic energy is indeed amplified by small-scale dynamo action~\citep{Graham2010}. We note that similar results are obtained for the simulation $\mathrm{d}h8$, although they are not displayed here for conciseness.
    
    In Table~\ref{tab:1mG}, we also report the time- and horizontally averaged unsigned vertical magnetic field and magnetic-to-kinetic energy density ratio, obtained during the saturation phase at $z=0$. For the case of $h=12\ \mathrm{km}$, for which the magnetic Reynolds number probably does not exceed $Re_{\mathrm{m},\mathrm{c}}$ enough, the magnetic energy and the unsigned vertical magnetic field settle to rather low values. On the other hand, the temporally and horizontally averaged unsigned vertical magnetic field reaches $49\ \mathrm{G}$ for $h=6\ \mathrm{km}$, with a magnetic to kinetic energy ratio of about $1.4\%$. We remind the reader that, while these values are about a factor of 1.5-3 smaller than the magnetic field strengths observed on the quiet Sun~\citep[see, e.g.][]{Trujillo2004,OrozcoSuarez2012}, the overall efficiency of our dynamo is strongly affected by the details of the numerical setup used in these simulations. As a matter of fact, we repeated the simulation $\mathrm{d}h8$ with $C_{\mathrm{lin}}=C_{\mathrm{sqrt}}=0$, obtaining $\tau_\mathrm{E}=1200\ \mathrm{s}$ and $\langle|B_z(z=0)|\rangle=30\ \mathrm{G}$. More importantly, the bottom boundary condition used in these simulations is very conservative, since we assumed a zero magnetic field in inflow regions. As also discussed in~\citet{Rempel2014}, this means that all magnetic field that is continuously pumped out of the computational domain across the bottom boundary is lost for the dynamo, and it is not replenished by magnetic field carrying upflows.
    
    In Sect.~\ref{sec:res_etanu}, we see that it is possible to approximately replicate the time-averaged vertical $Re_\mathrm{eff}$ and $Re_{\mathrm{m},\mathrm{eff}}$ profiles of simulation $h12$ when using $C_\mathrm{S}=1.55$ and $\eta_\mathrm{H}=3.5\cdot10^{10}\ \mathrm{cm^2\,s^{-1}}$ for the simulation with $h=6\ \mathrm{km}$. Therefore, we now investigate whether simulations of similar $Re_\mathrm{eff}$ and $Re_{\mathrm{m},\mathrm{eff}}$ produce similar e-folding times. To this aim, we carried out three additional simulations, always starting from a seed magnetic field of $1\ \mathrm{mG}$, with $C_\mathrm{S}=1.55$ and $\eta_\mathrm{H}=2.2,\ 2.5$, and $3.2\cdot10^{10}\ \mathrm{cm^2\,s^{-1}}$ (simulations d$h6\nu_\mathrm{S}\eta_\mathrm{H}22$, d$h6\nu_\mathrm{S}\eta_\mathrm{H}25$, and d$h6\nu_\mathrm{S}\eta_\mathrm{H}32$, respectively). From a study of the time evolution of the total magnetic energy in the boxes, it results that with $\eta_\mathrm{H}=3.2\cdot10^{10}\ \mathrm{cm^2\,s^{-1}}$ or larger, it is not possible to observe dynamo action (in Fig.~\ref{Fig:Em_t}, left, we show only the time-evolution profile of simulation d$h6\nu_\mathrm{S}\eta_\mathrm{H}25$ for clarity). On the other hand, the magnetic energy grows exponentially when $\eta_\mathrm{H}=2.5\cdot10^{10}\ \mathrm{cm^2\,s^{-1}}$ or less, corresponding to $Re_{\mathrm{m},\mathrm{eff}}=860$ or larger, although we could not afford to run until saturation, because of the extremely large computational cost of these simulations. Comparing the results in Table~\ref{tab:1mG} to simulation d$h12$, we find that, when using explicit viscosities and magnetic diffusivities, we need larger effective magnetic Reynolds numbers with respect to the case of using only implicit numerical dissipations to obtain dynamo action. To further analyse this kind of hysteresis, Fig.~\ref{Fig:TF} also displays the imbalance $-\partial_tE_\mathrm{m}+T_\mathrm{ms}+T_\mathrm{ma}$, with $\partial_tE_\mathrm{m}(k_\mathrm{h})=\hat{\vec{B}}(k_\mathrm{h})\cdot\widehat{\partial_t\vec{B}}^*(k_\mathrm{h})/(8\pi)+\mathrm{c.c.}$, indicated with light blue dotted curves, as an estimate of the numerical dissipation, as well as the energy transfer rates for simulation d$h6\nu_\mathrm{S}\eta_\mathrm{H}25$ (right panel). Comparing the results of the simulations d$h12$ and d$h6\nu_\mathrm{S}\eta_\mathrm{H}25$, we see that the profiles of $T_\mathrm{ms}$, $T_\mathrm{ma}$, and $T_\mathrm{kt}$ are rather similar, the main differences being at very small scales. On the other hand, while for the simulations d$h12$ and d$h6$ the term $T_\mathrm{md}$ overestimates $-\partial_tE_\mathrm{m}+T_\mathrm{ms}+T_\mathrm{ma}$ at large scales, but much better agrees at smaller scales (purple versus light blue curves), we see a much better agreement between the purple dashed and light blue dotted curves for the simulation d$h6\nu_\mathrm{S}\eta_\mathrm{H}25$. This implies that, to obtain an equal growth rate, it is not enough to have the same effective Reynolds and magnetic Reynolds numbers, it is also important that the dissipation terms operate at the same length scales.
    
    Finally, the simulations d$h6\eta_\mathrm{H}12$ and d$h6\eta_\mathrm{H}14$ were carried out to investigate the possibility of amplifying magnetic energy through dynamo action at smaller magnetic Prandtl numbers than 0.9. The corresponding magnetic energy e-folding times are reported in Table~\ref{tab:1mG}, while the time evolution of the magnetic energy resulting from the simulation d$h6\eta_\mathrm{H}12$ is displayed in Fig.~\ref{Fig:Em_t}. The results suggest that the magnetic Reynolds numbers of these two simulations are very close to, but larger than, $Re_{\mathrm{m},\mathrm{c}}$, and that the small-scale dynamo operates even at $Pr_{\mathrm{m},\mathrm{eff}}\simeq0.65$. Comparing these results to the critical magnetic Reynolds numbers of $Re_\mathrm{m,c}\simeq200-300$ obtained by~\citet{Schekochihin2005} for $Re\simeq1000-2000$, we note that CO\textsuperscript{5}BOLD simulations have about a factor four larger $Re_\mathrm{m,c}$. This means that we need a correspondingly smaller magnetic diffusivity to obtain dynamo action in these simulations.

\section{Conclusions}\label{sec:conclusions}
    In the present paper, we propose a methodology, based on the method of PoPe of~\citet{Cartier-Michaud2016}, for estimating the effective magnetic Prandtl number stemming from radiative MHD simulations. The methodology is simple, general, and can be applied as a post-processing step after having obtained the simulation results, without the need for modifying the simulation code or knowing the exact details of the employed numerical scheme. It just requires introducing a different numerical scheme than the one used in the simulation code (in particular it should be of higher order of accuracy),  recomputing the operators of the momentum and induction equations with the new scheme, and then minimising a residual through classical residual methods.
    
    The application of the proposed methodology to a number of CO\textsuperscript{5}BOLD simulations allowed us to investigate the advantages and drawbacks of the procedure. In particular, it is found that the results are almost insensitive to the order of accuracy of the post-processing scheme we tested. Also, the results are insensitive to the exact form used to express the stress tensor. Moreover, it is shown that it is also possible to obtain dissipation coefficients that depend on height, that the resulting effective dissipations are in rather good agreement with explicit diffusion coefficients if these are large enough, and that our results are consistent with an intrinsic viscosity acting approximately at the grid scale. Overall, we are now confident that the proposed methodology provides a solid estimate of the dissipation coefficients affecting the momentum and induction equations of MHD simulation codes and, consequently, a reliable evaluation of the magnetic Prandtl number characterising the numerical results. It also emerged that having the same effective dissipation coefficients does not ensure we obtain the same magnetic and kinetic energy spectra, as implicit and explicit dissipations might operate at different scales.
    
    Finally, the possibility of simulating small-scale dynamo action with CO\textsuperscript{5}BOLD was investigated. It is found that, for resolutions higher than or equal to $h=12\ \mathrm{km}$, small-scale dynamos are active and can amplify a small seed magnetic field up to significant values. However, as also found by other authors in the past, the small-scale dynamo strongly depends on the details of the numerical setup, and in particular of the boundary conditions used at the bottom of the numerical boxes and the exact nature of the magnetic diffusivity. It was also possible to identify dynamo action at rather low magnetic Prandtl numbers ($Pr_{\mathrm{m},\mathrm{eff}}\simeq0.65$), although these values are still far from the $Pr_\mathrm{m}\ll1$ regime characterising the solar atmosphere.

\begin{acknowledgements}
      This work was supported by the Swiss National Science Foundation under grant ID 200020\_182094. Part of the numerical simulations were carried out on Piz Daint at CSCS under project IDs sm51, s1059, and u14, and the rest was carried out on the HPC ICS cluster at USI. We sincerely thank the anonymous referee for the careful reading of the manuscript and the many comments, which greatly helped us to improve the quality of the paper.
\end{acknowledgements}


\bibliographystyle{aa} 
\bibliography{42644corr}

\end{document}